"A re-examination of Maxwell's electromagnetic equations"

J. Dunning-Davies,
Department of Physics,
University of Hull,
Hull,
England

*J.Dunning-Davies@hull.ac.uk*

**Abstract.**
   It is pointed out that the usual derivation of the well-known Maxwell electromagnetic equations holds only for a medium at rest. A way in which the equations may be modified for the case when the mean flow of the medium is steady and uniform is proposed. The implication of this for the problem of the origin of planetary magnetic fields is discussed.

**Introduction.**

Maxwell's electromagnetic equations are surely among the best known and most widely used sets of equations in physics. However, possibly because of this and since they have been used so successfully in so many areas for so many years, they are, to some extent, taken for granted and used with little or no critical examination of their range of validity. This is particularly true of the two equations

$$\nabla \times \mathbf{E} = -\frac{1}{c}\frac{\partial \mathbf{B}}{\partial t}$$

and

$$\nabla \times \mathbf{H} = 4\pi \mathbf{j} + \frac{1}{c}\frac{\partial \mathbf{D}}{\partial t}$$

Both these equations are used widely but, although the point is made quite clearly in most elementary, as well as more advanced, textbooks, it is often forgotten that these equations apply *only* when the medium involved is assumed to be at rest. This assumption is actually crucial in the derivation of these equations since it is because of it that it is allowable to take the operator d/d*t* inside the integral sign as a partial derivative and so finally derive each of the above equations. This leaves open the question of what happens if the medium is not at rest?

As is well known, for a non-conducting medium at rest, Maxwell's electromagnetic equations, when no charge is present, reduce to

$$\nabla \cdot \mathbf{E} = 0, \quad \nabla \times \mathbf{E} = -\frac{\mu}{c}\frac{\partial \mathbf{H}}{\partial t},$$

$$\nabla \cdot \mathbf{H} = 0, \quad \nabla \times \mathbf{H} = \frac{\varepsilon}{c}\frac{\partial \mathbf{E}}{\partial t},$$

where $\mathbf{D} = \varepsilon \mathbf{E}, \mathbf{B} = \mu \mathbf{H}$ and $\mu, \varepsilon$ are assumed constant in time.

The first two equations are easily seen to lead to

$$\nabla^2 \mathbf{E} = \frac{\varepsilon\mu}{c^2} \frac{\partial^2 \mathbf{E}}{\partial t^2},$$

and the latter two to

$$\nabla^2 \mathbf{H} = \frac{\varepsilon\mu}{c^2} \frac{\partial^2 \mathbf{H}}{\partial t^2}.$$

Therefore, in this special case, *provided* the medium is at rest, both E and H satisfy the well-known wave equation. However, it has been shown [Thornhill, 1993] that, if the mean flow is steady and uniform, and, therefore, both homentropic and irrotational, the system of equations governing small-amplitude homentropic irrotational wave motion in such a flow reduces to the equation

$$\nabla^2 \phi = (1/c^2) D^2 \phi / Dt^2,$$

which is sometimes referred to as the convected, or progressive, wave equation. The question which remains is, for the case of a medium not at rest, should Maxwell's electromagnetic equations be modified so as to reduce to this progressive wave equation in the case of a non-conducting medium with no charge present?

**Generalisation of Maxwell's equations.**

In the derivation of

$$\nabla \times \mathbf{E} = -\frac{\mu}{c} \frac{\partial \mathbf{H}}{\partial t}$$

it proves necessary to consider the integral

$$-\frac{\mu}{c} \frac{d}{dt} \int \mathbf{B}.\mathbf{dS}$$

and, as stated previously, interchange the derivative and the integral. This operation may be carried out only for a medium at rest. However, if the medium is moving, then the surface $S$ in the integral will be moving also, and the mere change of $S$ in the field $\mathbf{B}$ will cause changes in the flux. Hence, following Abraham and Becker [1932], a new kind of differentiation with respect to time is defined by the symbol $\dot{\mathbf{B}}$ as follows:

$$\frac{d}{dt}\int \mathbf{B}.\mathbf{dS} = \int \dot{\mathbf{B}}.\mathbf{dS} \qquad (a)$$

Here, $\dot{\mathbf{B}}$ is a vector, the flux of which across the moving surface equals the rate of increase with time of the flux of $\mathbf{B}$ across the same surface. In order to find $\dot{\mathbf{B}}$, the exact details of the motion of the surface concerned must be known. Suppose this motion described by a vector $\mathbf{u}$, which is assumed given for each element $dS$ of the surface and is the velocity of the element.

Let $S_1$ be the position of the surface $S$ at time $(t-dt)$ and $S_2$ the position at some later time $t$. $S_2$ may be obtained from $S_1$ by giving each element of $S_1$ a displacement $\mathbf{u}dt$. The surfaces $S_1$ and $S_2$, together with the strip produced during the motion, bound a volume $dt\int \mathbf{u}.\mathbf{dS}$.

The rate of change with time of the flux of $\mathbf{B}$ across $S$ may be found from the difference between the flux across $S_2$ at time $t$ and that across $S_1$ at time $(t-dt)$; that is

$$\frac{d}{dt}\int \mathbf{B}.\mathbf{dS} = \frac{\int \mathbf{B}_t.\mathbf{dS}_2 - \int \mathbf{B}_{t-dt}.\mathbf{dS}_1}{dt},$$

where the subscript indicates the time at which the flux is measured.

The divergence theorem may be applied at time $t$ to the volume bounded by $S_1$, $S_2$ and the strip connecting them. Here the required normal to $S_2$ will be the outward pointing normal and that to $S_1$ the inward pointing normal. Also, a surface element of the side face will be given by $\mathbf{ds}\times \mathbf{u}dt$. Then, the divergence theorem gives

$$\int_{S_2} \mathbf{B}_t.\mathbf{dS}_2 + dt\oint \mathbf{B}.\mathbf{ds}\times \mathbf{u} - \int_{S_1} \mathbf{B}_t.\mathbf{dS}_1 = dt\int (\nabla.\mathbf{B})\mathbf{u}.\mathbf{dS}.$$

Also

$$\int \mathbf{B}_{t-dt} \cdot \mathbf{dS}_1 = \int \mathbf{B}_t \cdot \mathbf{dS}_1 - \int \frac{\partial \mathbf{B}}{\partial t} \mathbf{dS}_1 dt .$$

Hence,

$$\int \mathbf{B}_t \cdot \mathbf{dS}_2 - \int \mathbf{B}_{t-dt} \cdot \mathbf{dS}_1 = dt \left\{ \int \dot{\mathbf{B}} \cdot \mathbf{dS}_1 + \int (\nabla \cdot \mathbf{B}) \mathbf{u} \cdot \mathbf{dS}_1 - \oint \mathbf{B} \cdot \mathbf{ds} \times \mathbf{u} \right\}.$$

Using Stokes' theorem, the final term on the right-hand side of this equation may be written

$$\oint \mathbf{B} \cdot \mathbf{ds} \times \mathbf{u} = \oint \mathbf{u} \times \mathbf{B} \cdot \mathbf{ds} = \int \{\nabla \times (\mathbf{u} \times \mathbf{B})\} \cdot \mathbf{dS},$$

so that finally

$$\frac{d}{dt} \int \mathbf{B} \cdot \mathbf{dS} = \int \left\{ \frac{\partial \mathbf{B}}{\partial t} + \mathbf{u}(\nabla \cdot \mathbf{B}) - \nabla \times (\mathbf{u} \times \mathbf{B}) \right\} \cdot \mathbf{dS}.$$

Therefore, the $\dot{\mathbf{B}}$, introduced in equation (a) above, is given by

$$\dot{\mathbf{B}} = \frac{\partial \mathbf{B}}{\partial t} + \mathbf{u}(\nabla \cdot \mathbf{B}) - \nabla \times (\mathbf{u} \times \mathbf{B})$$

or, noting that

$$\nabla \times (\mathbf{u} \times \mathbf{B}) = \mathbf{u}(\nabla \cdot \mathbf{B}) - \mathbf{B}(\nabla \cdot \mathbf{u}) + (\mathbf{B} \cdot \nabla)\mathbf{u} - (\mathbf{u} \cdot \nabla)\mathbf{B} ,$$

$$\dot{\mathbf{B}} = \frac{\partial \mathbf{B}}{\partial t} + (\mathbf{u} \cdot \nabla)\mathbf{B} + \mathbf{B}(\nabla \cdot \mathbf{u}) - (\mathbf{B} \cdot \nabla)\mathbf{u}$$

However, if the mean flow is steady and uniform and     , therefore, both homentropic and irrotational, the fluid velocity, **u,** will be constant and this latter equation will reduce to

$$\dot{\mathbf{B}} = \frac{\partial \mathbf{B}}{\partial t} + (\mathbf{u} \cdot \nabla)\mathbf{B} = \frac{D\mathbf{B}}{Dt} ,$$

that is, for such flow, $\dot{\mathbf{B}}$ becomes the well-known Euler derivative. It might be noted, though, that, for more general flows, the expression for $\dot{\mathbf{B}}$ is somewhat more complicated.

It follows that, if the mean flow is steady and uniform, the Maxwell equation, mentioned above, becomes

$$\nabla \times \mathbf{E} = -\frac{\mu}{c}\frac{D\mathbf{H}}{Dt} = -\frac{\mu}{c}\left[\frac{\partial \mathbf{H}}{\partial t} + (\mathbf{u}.\nabla)\mathbf{H}\right]..$$

Also, in this particular case, the remaining three Maxwell equations will be

$$\nabla.\mathbf{E} = 0, \quad \nabla.\mathbf{H} = 0,$$

$$\nabla \times \mathbf{H} = \frac{\varepsilon}{c}\frac{D\mathbf{E}}{Dt} = \frac{\varepsilon}{c}\left[\frac{\partial \mathbf{E}}{\partial t} + (\mathbf{u}.\nabla)\mathbf{E}\right],$$

with this form for the final equation following in a manner similar to that adopted above when noting that, for a steady, uniform mean flow, $\partial/\partial t$ is replaced by $D/Dt$ in the equation for $\nabla \times \mathbf{E}$.

These four modified Maxwell equations lead to both $\mathbf{E}$ and $\mathbf{H}$ satisfying the above mentioned progressive wave equation, as they surely must.

**The origin of planetary magnetic fields.**

It is conceivable that use of these modified Maxwell electromagnetic equations could provide new insight into the problem of the origin of planetary magnetic fields. This is a problem which has existed, without a really satisfactory explanation, for many years. It would seem reasonable to expect all such fields to arise from the same physical mechanism, although the minute detail might vary from case to case. The mechanism generally favoured as providing the best explanation for the origin of these fields was the dynamo mechanism, although the main reason for its adoption was the failure of the alternatives to provide a consistent explanation. However, Cowling [1934] showed that there is a limit to the degree of symmetry encountered in a steady dynamo mechanism; this result, based on the traditional electromagnetic equations of Maxwell, shows that the steady maintenance of a poloidal field is simply not possible - the result is in reality an anti-dynamo theorem which raises difficulties in understanding the observed symmetry of the dipole field.

Following Alfvén [1963], it might be noted that, in a stationary state, there is no electromagnetic field along a neutral line because that would imply a non-vanishing $\nabla \times \mathbf{E}$, and so a time varying $\mathbf{B}$. The induced electric field $\mathbf{v} \times \mathbf{B}$ vanishes on the neutral line since $\mathbf{B}$ does. Thus, there can be no electromotive force along the neutral line, and therefore the current density in the stationary state vanishes, the conductivity being infinite. On the other hand, $\nabla \times \mathbf{B}$ does not vanish on the neutral line. By Maxwell's usual equations, the non-vanishing $\nabla \times \mathbf{B}$ and the vanishing current density are in contradiction and so the existence of a rotationally symmetric steady-state dynamo is disproved. However, this conclusion may not be drawn if the modified Maxwell equations, alluded to earlier, are used, since, even in the steady state where the partial derivatives with respect to time will all be zero, the equation for $\nabla \times \mathbf{B}$ will reduce to

$$\nabla \times \mathbf{B} = \frac{1}{\mu}\left[\mathbf{j} + \varepsilon\frac{\partial \mathbf{E}}{\partial t} + \varepsilon \mathbf{v}.\nabla\mathbf{E}\right] \to \frac{\varepsilon}{\mu}\mathbf{v}.\nabla\mathbf{E}$$

and there is no reason why this extra term on the right-hand side should be identically equal to zero. Also, the non-vanishing of $\nabla \times \mathbf{E}$ will not imply a time varying $\mathbf{B}$ since, once again, there is an extra term - $\mathbf{v}.\nabla\mathbf{B}$ remaining to equate with the $\nabla \times \mathbf{E}$. It follows that an electromagnetic field may exist along the neutral line under these circumstances. Hence, no contradiction occurs; instead, a consistent system of differential equations remains to be solved.